\documentclass[twocolumn,prl,aps,showpacs]{revtex4}
\usepackage{epsfig,latexsym,graphicx}

\def \ba {\begin{eqnarray}}
\def \ea {\end{eqnarray}}

\begin{document}

\title{ Model for the inverse isotope effect of FeAs-based superconductors
in the $\pi$-phase-shifted pairing state }

\author{Han-Yong Choi\email[To whom the
correspondences should be addressed:~]{ hychoi@skku.edu}}
\affiliation{Department of Physics and Institute for Basic
Science Research, SungKyunKwan University, Suwon 440-746, Korea.}

\author{Jae Hyun Yun}
\affiliation{Department of Physics and Institute for Basic
Science Research, SungKyunKwan University, Suwon 440-746, Korea.}

\author{Yunkyu Bang}\affiliation{Department of Physics, Chonnam National
University, Kwangju 500-757, Korea. \\
Asia Pacific Center for Theoretical Physics, Pohang 790-784,
Korea.}

\author{Hyun C. Lee}\affiliation{Department of Physics and Basic Science Research
Institute, Sogang University, Seoul 121-742, Korea}

\begin{abstract}

The isotope effects for Fe based superconductors are considered by
including the phonon and magnetic fluctuations within the two band
Eliashberg theory. We show that the recently observed inverse
isotope effects of Fe, $\alpha_{Fe} \approx -0.18 \pm 0.03
$,\cite{Shirage0903.3515} as well as the large positive isotope
exponent ($\alpha\approx 0.35$) can naturally arise for the
magnetically induced sign revered $s$-wave pairing state within
reasonable parameter range. Either experimental report can not be
discarded from the present analysis based on the parameter values
they require. The inverse and positive isotope effects mean,
respectively, the interband and intraband dominant eletron-phonon
interaction. We first make our points based on the analytic
result from the square well potential model and present explicit
numerical calculations of the two band Eliashberg theory.

\end{abstract}



\maketitle

$Introduction$ -- The alluring prospect of opening a key window to
understanding the mechanism of high temperature superconductivity
has attracted fierce research activities in the iron based
pnictides.\cite{Kamihara08jacs,ChenGF08prl,ChenXH08nature} The
first step towards this path is to establish the pairing symmetry
and dynamics of the FeAs superconductors. Among the many ideas put
forward, particularly appealing is the sign reversed pairing
state.\cite{Mazin08prl,Kuroki08prl} It is the ground state of a
two band superconductivity where both pairing order parameters,
$\Delta_1$ and $\Delta_2$, on the two bands have full gaps while
acquiring the $\pi$ phase shift between them. A repulsive
interband interaction is turned to induce pairing by generating
the sign reversal between the two order parameters. The
subsequent theoretical studies supported the idea of the sign
reversed full gaps. The $s_\pm$ or $s\pi$ pairing state seems to
be able explain the experimental observations indicating the full
gap as well as a gap with
nodes.\cite{Bang08prb,Chubukov08prb,Raghu08prb,Choi0807.4604} The
evidence for the full gaps came from the several independent
ARPES experiments on single crystals
\cite{Ding08epl,Zhao0807.0398,Kondo0807.0815}. The phase shift of
$\pi$ may account for the absence of the Hebel-Slichter peak and
the low temperature behavior of $\sim T^3$ in the spin-lattice
relaxation rate
$1/T_1$,\cite{Matano0806.0249,Grafe0805.2595,Nakai0804.4765}
which seemed difficult to understand in terms of a full gap
scenario.

As for the pairing interaction, one widespread school of thought
maintains that magnetic fluctuations are intimately involved for
the superconductivity. The superconductivity emerges out of the
interplay between the density of charge carriers and strength of
antiferromagnetic (AF) interaction such that their combined effect
becomes optimum. This view seems natural from the overall phase
diagram of the pnictides in the temperature and doping plane. The
supercondcutivity emerges from the parent metallic
antiferromagnetic state as the static AF order is suppressed and
charge carriers are introduced through the chemical
doping.\cite{Fang0903.2418} This view is further supported by the
neutron scattering investigations showing a resonance at a wave
vector related to the AF order coincident with the onset of
superconductivity.\cite{Christianson08nature} Within this school
of thought combined with the theoretical estimates of the small
electron-phonon coupling constant, it came rather as a surprise
that a large isotope coefficient $\alpha \approx 0.34-0.37$ was
observed by Liu $et~al$. with a substitution of $^{56}$Fe by
$^{54}$Fe in SmFeAsO$_{1-x}$F$_x$ ($x=0.15$) and
Ba$_{1-x}$K$_x$Fe$_2$As$_2$ ($x=0.4$).\cite{Liu0810.2694}
Theoretical calculation by Boeri $et~al.$ reported a very weak
electron-phonon coupling constant ($\lambda_{ ph} \approx
0.21$),\cite{Boeri08prl} which is too small to account for the
observed $T_c$ of the pnictides. Recall $\alpha$ is indeed very
small for the cuprate supercondcutors particularly near the
optimal doping.\cite{Franck94review} Perhaps even more surprising
is the observation of the sizable $inverse$ isotope effect of
$\alpha_{Fe} = -0.18 \pm0.03$ by Shirage $et~al.$ in
Ba$_{1-x}$K$_x$Fe$_2$As$_2$ ($x=0.4$)\cite{Shirage0903.3515}

In view of this controversy, it is important that the Liu $et~al.$
and Shirage $et~ al.$ results be confirmed. What we study here is
to check if these experimental results are compatible with the
current understanding of the pnictide superconductivity. We can
estimate the parameters like the coupling constants that these
experiments require. These motivate us to revisit the issue of
electron-phonon coupling in the Fe pnictides. We argue that the
inverse isotope effect as well as the large positive effect can
arise with a reasonable parameter range within the widely held
view of the magnetically mediated $\pi$ phase shifted $s$-wave
pairing scheme. We first make our points based on the analytic
result from the square well potential model applied to the sign
changing $s$-wave state, and substantiate them by the explicit
numerical calculations of the two band Eliashberg theory.

{\it Idea and Model} -- As sketched below in the $Formalism$
section, a straightforward application of the square well
potential model to the case where there are AF spin fluctuations
and phonons yields that the critical temperature $T_c$ and
isotope exponent $\alpha$ are given as follows:
 \ba\label{tc}
T_c &=& 1.14 \omega_{ph}
\exp\left(-\frac{1+\lambda_{AF}^{+}+\lambda_{ph}^{+}}
{\Lambda_{eff}-\lambda_{ph}^{-}}\right),
 \\
 \label{alpha}
\alpha &=& \frac{1}{2} \left[1-
\frac{1+\lambda_{AF}^{+}+\lambda_{ph}^{+}} {1+\lambda_{AF}^{+}}
\left( \frac{\Lambda_{eff}}{\Lambda_{eff}-\lambda_{ph}^{-}}
\right)^2 \right],
 \\ \label{lambdaeff} \Lambda_{eff} &=&\frac{\lambda_{AF}^{-}}
 {1- \frac{\lambda_{AF}^{-}}{1+\lambda_{AF}^{+}}
\ln\left(\frac{\omega_{AF}}{\omega_{ph}} \right) },
 \ea
where $\omega_{AF}$ and $\omega_{ph}$ are the AF fluctuation and
phonon frequencies, and $\lambda_{AF/ph}^\pm$ is given by the
sum/difference between the interband and intraband interactions as
 \ba
 \label{lambda12}
\lambda_{AF/ph}^\pm = \lambda_{AF/ph}^{(inter)} \pm
\lambda_{AF/ph}^{(intra)}.
 \ea
Note that the $T_c$ formula of Eq.\ (\ref{tc}) is reduced to the
formula in Shirage $et~al$.\cite{Shirage0903.3515} in the limit
that $\lambda^{(intra)}=0$ and the quasi-particle renormalization
is neglected. Also similar formulas were obtained before in
different contexts.\cite{Shimahara03jpsj,Dolgov05prl} The result
of Eqs.\ (\ref{tc}) and (\ref{alpha}), which will be backed up
below by numerically solving the two band Eliashberg theory,
explicitly shows that depending on whether $\lambda_{ph}^{-}
\gtrsim 0$ or $ \lesssim 0$, the isotope effect can show the
inverse or conventional behavior.

Two factors are behind this interesting effect: the $-$ sign
between the $\Lambda_{eff}$ and $\lambda_{ph}^-$ of Eq.\
(\ref{alpha}) and the possibility of both signs of
$\lambda_{ph}^-$ of Eq.\ (\ref{lambda12}). The first factor comes
from that the phonon and spin fluctuations enter the pairing
kernel $\lambda_{ij}^{(-)}$ with the opposite sign as in Eqs.\
(\ref{deltapn}) and (\ref{lambdapm}), and the second factor from
the sign change between the two gaps as suggested by the $s_\pm$
model.\cite{Mazin08prl,Kuroki08prl} An inspection of $\alpha$ of
Eq.\ (\ref{alpha}) then reveals that the inverse isotope effect
implies that the interband electron-phonon coupling is comparable
with or stronger than the intraband coupling. Note also that
$\lambda_{ph}^{inter}=\lambda_{ph}^{intra}$ gives the inverse
isotope effect except when both are equal to zero. Also
noteworthy is that when $\Lambda_{eff}- \lambda_{ph}^{-} \approx
0$, $T_c$ becomes very small and a giant isotope effect will
appear, $|\alpha| \gg 1/2$. It seems that both the large isotope
effect and the inverse isotope effect can arise in the
magnetically induced $\pi$ phase shifted pairing for the
pnictides within a reasonable parameter range. Either
experimental report can not be discarded from the present
analysis based on the parameter values they require.

Although this discussion demonstrates the essential physics, it
neglects the more general spin and phonon interactions and
involves the square well potential model which is not a controled
approximation. We will therefore present the results of the
numerical calculations to show that inclusion of these effects
still support this conclusion. We will show this by two band
Eliashberg calculations as presented in what follows.

$Formalism$ -- The two-band Eliashberg equation with an isotropic
gap in each band may be written in the Matsubara frequency as
 \ba
 \label{zpn}
Z_i(ip_n) = 1 +\frac{\pi T}{p_n} \sum_{j,m}
\lambda_{ij}^{(+)}(i\omega_n) \frac{p_m}{\sqrt{p_m^2 +\Delta_j
(ip_m)^2}}, \\
 \label{deltapn}
\Delta_i (ip_n) Z_i (ip_n) = \pi T \sum_{j,m}
\lambda_{ij}^{(-)}(i\omega_n) \frac{\Delta_j (ip_m)}{\sqrt{p_m^2
+\Delta_j(ip_m)^2}} ,
 \ea
where $i,j=1,2$ are the band indices, $\omega_n=p_n-p_m$, $T$ the
temperature, and $p_n$ and $p_m$ are the Matsubara frequencies.
The kernel $\lambda_{ij}^{(\pm)}$ are given by
 \ba
 \label{lambdapm}
\lambda_{ij}^{(\pm)}(i\omega_n) = \lambda_{ij}^{(ph)}(i\omega_n)
\pm \lambda_{ij}^{(AF)}(i\omega_n) .
 \ea
We take the phonon and spin fluctuations as
 \ba
\lambda_{ij}^{(AF/ph)}(i\omega_n) &=& \lambda^{AF/ph}_{ij}
\int_0^\infty d\epsilon \frac{\omega_{\nu} \epsilon} {\epsilon^2
+\omega_n^2} F(\epsilon) ,
 \ea
where $F(\epsilon)$ is a truncated Lorentzian centered at
$\omega_{\nu}=\omega_{AF/ph}$
 \ba
F(\epsilon) = \left\{ \begin{array}{ll} \frac{1}{R} \left[
\frac{1}{(\epsilon-\omega_{\nu} )^2 +\Gamma^2}
-\frac{1}{\Gamma_c^2 +\Gamma^2} \right], & {\rm for} ~|\epsilon
-\omega_{\nu} | \le \Gamma_c , \\
0, & {\rm otherwise} ,
\end{array} \right.
 \ea
We took $\Gamma_c=\omega_\nu$, $\Gamma=\Gamma_c/2$, and $R$ is a
normalization constant such that $\int_0^{\infty} d\epsilon F
(\epsilon) = 1$.\cite{Bickers00prb} The $\Gamma_c=\omega_\nu$ was
chosen such that $F(\epsilon)\sim\omega$ for small $\omega$ to
simulate the overdamped AF spin fluctuations.

Eqs.\ (\ref{zpn}) and (\ref{deltapn}) may be linearized near
$T=T_c$, and may be written as the following matrix form in the
basis of $\Delta_i(ip_n)$.
 \ba
 \label{lineardelta}
\Delta_i (ip_n) = \sum_j \sum_m M_{ij}(n,m) \Delta_j (ip_m) ,
 \ea
where the matrix $M$ is defined as
 \ba
 \label{linearz}
M_{ij}(n,m) &=& \frac{\pi T}{|p_m|} \frac{\lambda_{ij}^{(-)}(ip_n-ip_m)}{Z_i(ip_n)}, \nonumber \\
 Z_i(ip_n) &=& 1 +\frac{\pi T}{p_n} \sum_{j,m}
\lambda_{ij}^{(+)}(ip_n-ip_m) sgn(p_m).
 \ea
The $T_c$ is then determined to be the temperature at which the
largest eigenvalue of the matrix $M$ becomes 1. The isotope
coefficient is given by
 \ba
\alpha = \frac12 \frac{\partial \ln T_c}{\partial \ln
\omega_{ph}} .
 \ea

$Preliminary ~Analysis$ -- Let us first consider some preliminary
analytic analysis based on the square well potential model
applied to the $s_\pm$ state. After neglecting the frequency
dependence of $\lambda_{ij}^{(\pm)} (i\omega_n)$ and taking the
standard approximation of $\Delta_i(\omega) =\Delta_{ia}$ for
$0<\omega<\omega_{ph}$ and $\Delta_{ib}$ for
$\omega_{ph}<\omega<\omega_{AF}$, and similarly for $Z_i
(\omega)$, we can rewrite Eqs.\ (\ref{linearz}) and
(\ref{lineardelta}) as
 \ba
Z_{ia} &=& 1+ \sum_j \lambda_{ij}^{(+)}, \\
Z_{ib} &=& 1+ \sum_j \lambda_{ij}^{AF}, \\
\Delta_{ia} Z_{ia} &=& \sum_j \left[ \lambda_{ij}^{(-)} F_a
\Delta_{ja} -\lambda_{ij}^{AF} F_b \Delta_{jb}
\right], \\
 \Delta_{ib} Z_{ib} &=& -\sum_j \left[
\lambda_{ij}^{AF} F_a \Delta_{ja} + \lambda_{ij}^{AF} F_b
\Delta_{jb} \right],
 \ea
where the band indices take $i,j=1,2$, and
 \ba
F_a &=& \int_0^{\omega_{ph}} d\xi \frac{1}{\xi}
\tanh\left(\frac{\xi}{2T_c} \right) = \ln\left( \frac{1.14
\omega_{ph}}{T_c} \right), \nonumber \\
 F_b &=& \ln\left( \frac{
\omega_{AF}}{\omega_{ph}} \right).
 \ea
A straightforward manipulation taking advantage of the relations
$\Delta_{1a} = -\Delta_{2a}$ and $\Delta_{1b} = -\Delta_{2b}$ of
the $s_\pm$ state gives Eqs. (\ref{tc}), (\ref{alpha}), and
(\ref{lambdaeff}) given above. Here, we used the definitions
$\lambda_{AF/ph}^\pm = \lambda_{12}^{AF/ph} \pm
\lambda_{11}^{AF/ph}$ and considered the simple case of
$\lambda_{12}^{AF/ph}=\lambda_{21}^{AF/ph}$ and
$\lambda_{11}^{AF/ph}=\lambda_{22}^{AF/ph}$.

Before we present the results from the numerical computations of
the linearized Eliashberg equation of Eq.\ (\ref{lineardelta}),
we first consider the approximate solutions from Eqs.\ (\ref{tc})
and (\ref{alpha}) to get ourselves oriented to the relevant
parameter space. Let us first consider the inverse isotope effect
in the context of the $s\pi$ pairing state. This may be realized
when the interband dominates the intraband phonon interaction as
discussed above. We take the simple parameterization of the
interband interactions only, $\lambda_{AF/ph}^\pm =
\lambda_{AF/ph}$ ($
\lambda_{11}^{AF/ph}=\lambda_{22}^{AF/ph}=0$). From $T_c\approx
3.0$ meV and $\alpha\approx -0.2$ reported by Shirage $et~al.$ for
Ba$_{1-x}$K$_x$Fe$_2$As$_2$, we get $\lambda_{AF}=1.53$ and
$\lambda_{ph}=0.189$ for $\omega_{ph}=\omega_{AF}=20$ meV, and
$\lambda_{AF}=0.944$ and $\lambda_{ph}=0.146$ for
$\omega_{ph}=20$, $ \omega_{AF}=30$ meV. The obtained
$\lambda_{AF}>\lambda_{ph}$ is consistent with the magnetically
induced pairing.

We now turn to the $\alpha>0$ case, which one of the authors
considered previously.\cite{Bang09prb} This may be realized in
the context of the $s\pi$ pairing state by $\lambda_{AF}^\pm =
\lambda_{AF}$ and $\lambda_{ph}^\pm = \pm\lambda_{ph}$. From
$T_c\approx 3.5$ meV and $\alpha\approx 0.35$ reported by Liu
$et~al.$ for SmFeAsO$_{1-x}$F$_x$, we get $\lambda_{AF}=0.836$,
and $\lambda_{ph}=0.738$ by solving Eqs.\ (\ref{tc}) and
(\ref{alpha}) for $\omega_{ph}=20 $ meV and $ \omega_{AF}=10$
meV. The solution for $\omega_{ph}=\omega_{AF}=20$ meV does not
satisfy $\lambda_{AF}>\lambda_{ph}$ relevant for the magnetically
mediated superconductivity. The inverse or positive isotope
effect is naturally obtained when $\omega_{AF} > \omega_{ph}$ or
$<$. This was pointed out by Shimahara in a different
context.\cite{Shimahara03jpsj}

In the derivation of Eqs.\ (\ref{tc}) and (\ref{alpha}), we
assumed $\omega_{AF} > \omega_{ph}$; if $\omega_{AF} <
\omega_{ph}$, they are $not$ valid. Also they are based on the
uncontrolled double well approximation. The correctness of the
results is confirmed by numerical calculations we are presenting
below.

\begin{figure}[hbt]
\vspace{-0.5cm}\label{fig1}
 \epsfxsize=7cm \epsffile{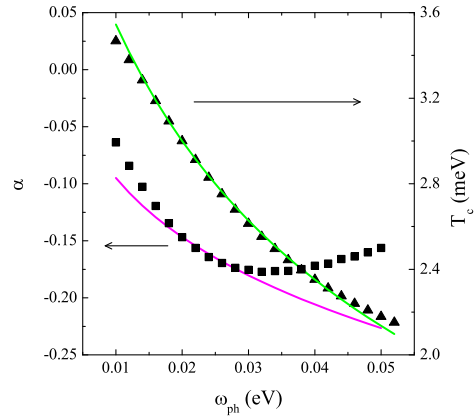}
\vspace{-0.5cm} \caption{The transition temperature $T_c$ and
isotope exponent $\alpha$ as a function of the phonon frequency
$\omega_{ph}$. $\omega_{AF}=30$ meV, $\lambda_{AF}=2.26$,
$\lambda_{ph}=0.5$, and all the intraband couplings are set equal
to 0. The solid lines are the plots of Eqs.\ (\ref{tc}) and
(\ref{alpha}). See the text for their parameters.}
\end{figure}

$Numerical~Results$ -- In Fig.\ 1, we show the results of
numerical calculations of the inverse isotope effect. We took,
guided by the above discussion, $\omega_{AF}=30$ meV, and adjusted
the interband $\lambda_{AF}$ and $\lambda_{ph}$ with all
intraband interactions set to 0, so that the $T_c$ and $\alpha$
give the right values for Ba$_{1-x}$K$_x$Fe$_2$As$_2$. With
$\lambda_{AF}=2.26$ and $\lambda_{ph}=0.5$, $\omega_{ph}$ was
varied between 10 and 50 meV. The solid squares and triangles
show the $\alpha$ and $T_c$, respectively. The lines are the
plots from Eqs.\ (\ref{tc}) and (\ref{alpha}).
$\lambda_{AF}=0.876$ and $\lambda_{ph}=0.104$ were selected such
that $T_c$ and $\alpha$ agree with the numerical calculations at
$\omega_{ph}=20$ and $\omega_{AF}=30$ meV. The interaction
parameters are somewhat different between the numerical and
(approximate) analytic calculations.

\begin{figure}[hbt]
\vspace{-0.5cm} \label{fig2}
 \epsfxsize=7cm \epsffile{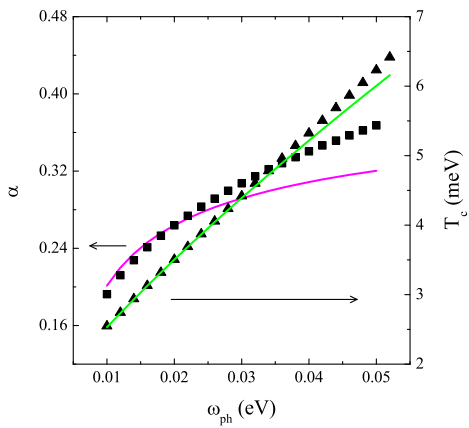}
\vspace{-0.5cm} \caption{$T_c$ and $\alpha$ as a function of
$\omega_{ph}$. $\omega_{AF}=15$ meV, $\lambda_{AF}=1.06$,
$\lambda_{ph}^{(intra)}=0.950$. The solid lines are the plots of
Eqs.\ (\ref{tc}) and (\ref{alpha}). See the text for their
parameters.}
\end{figure}

We then consider the positive isotope effect and present the
results in Fig.\ 2. We took $\omega_{AF}=15$ meV, interband
$\lambda_{AF}=1.06$, intraband $\lambda_{ph}=0.950$, and varied
$\omega_{ph}$ between 10 and 50 meV as before, with other
interactions set to 0. The solid triangles and squares show the
$\alpha$ and $T_c$, respectively, and the lines are from the
analytic expression with $\lambda_{AF}=0.878$ and
$\lambda_{ph}=0.491$ to agree with the numerical results at
$\omega_{AF}=15$ and $\omega_{ph}=20$ meV.

Although we have not fine-tuned the parameters to match the
observed $T_c$ and $\alpha$ exactly for the
Ba$_{1-x}$K$_x$Fe$_2$As$_2$ and SmFeAsO$_{1-x}$F$_x$, some
features are worth noting. First, the parameters for the numerical
and analytical calculations to give the observed $T_c$ and
$\alpha$ were somewhat different. This is because $\lambda_{AF}$
and $\lambda_{ph}$ vary sensitively as $\omega_{AF}$ is varied
(but not so sensitively on $\omega_{ph}$) in this region of
parameter space as one can see easily by solving Eqs.\ (\ref{tc})
and (\ref{alpha}). Second, it is necessary that the phonons are
predominantly of interband/intraband character for the
inverse/positive isotope effects. And, third, the phonon coupling
constant somewhat larger than the theoretical calculation seems
necessary.\cite{Boeri08prl} But, this might be understood by
including other effects like the different density of states
between the bands.\cite{Bang09prb}

$Remarks$ -- We have shown that both the inverse isotope effect
and large isotope exponent reported in the pnictides may arise in
the magnetically induced $\pi$ phase shifted pairing with
reasonable parameters. Either experimental report can not be
discarded from the present analysis based on the parameter values
they require. The dominant interband (intraband) phonon
interaction produces the inverse (positive) isotope effect. Note
that the contradicting inverse and positive isotope effects were
observed in the same compound, Ba$_{1-x}$K$_x$Fe$_2$As$_2$
($x=0.4$), by Fe substitution. The Liu result was obtained on the
sample from the solid state reaction whereas the Shirage result
from the high pressure technique. If either result turns out to be
reproducible, it will point to the interband or intraband
dominant electron-phonon interaction in the pnictides. Liu
$et~al$ also observed that O isotope effect is quite small,
$\alpha_O \approx -0.06$. In the framework proposed here, this
may be understood in terms of $\lambda_{ph}^+ \approx
\lambda_{ph}^- \approx 0$ for the O ions. It remains to be checked
if this anticipation is correct. It will be also interesting to
see if the spin density wave isotope exponent reported by Liu
$et~al$ may be understood within the proposed idea here.

$Acknowledgements$

This work was supported by Korea Research Foundation (KRF)
through Grant No.\ KRF-C00241.


\begin{thebibliography}{25}
\expandafter\ifx\csname
natexlab\endcsname\relax\def\natexlab#1{#1}\fi
\expandafter\ifx\csname bibnamefont\endcsname\relax
  \def\bibnamefont#1{#1}\fi
\expandafter\ifx\csname bibfnamefont\endcsname\relax
  \def\bibfnamefont#1{#1}\fi
\expandafter\ifx\csname citenamefont\endcsname\relax
  \def\citenamefont#1{#1}\fi
\expandafter\ifx\csname url\endcsname\relax
  \def\url#1{\texttt{#1}}\fi
\expandafter\ifx\csname
urlprefix\endcsname\relax\def\urlprefix{URL }\fi
\providecommand{\bibinfo}[2]{#2}
\providecommand{\eprint}[2][]{\url{#2}}

\bibitem[{\citenamefont{Shirage et~al.}(2009)\citenamefont{Shirage, Kihou,
  Miyazawa, Lee, Kito, Eisaki, Tanaka, and Iyo}}]{Shirage0903.3515}
\bibinfo{author}{\bibfnamefont{P.~M.} \bibnamefont{Shirage}},
  \bibinfo{author}{\bibfnamefont{K.}~\bibnamefont{Kihou}},
  \bibinfo{author}{\bibfnamefont{K.}~\bibnamefont{Miyazawa}},
  \bibinfo{author}{\bibfnamefont{C.-H.} \bibnamefont{Lee}},
  \bibinfo{author}{\bibfnamefont{H.}~\bibnamefont{Kito}},
  \bibinfo{author}{\bibfnamefont{H.}~\bibnamefont{Eisaki}},
  \bibinfo{author}{\bibfnamefont{Y.}~\bibnamefont{Tanaka}}, \bibnamefont{and}
  \bibinfo{author}{\bibfnamefont{A.}~\bibnamefont{Iyo}},
  \bibinfo{journal}{arXiv:0903.3515}  (\bibinfo{year}{2009}).

\bibitem[{\citenamefont{Kamihara et~al.}(2008)\citenamefont{Kamihara, Watanabe,
  Hirano, , and Hosono}}]{Kamihara08jacs}
\bibinfo{author}{\bibfnamefont{Y.}~\bibnamefont{Kamihara}},
  \bibinfo{author}{\bibfnamefont{T.}~\bibnamefont{Watanabe}},
  \bibinfo{author}{\bibfnamefont{M.}~\bibnamefont{Hirano}}, , \bibnamefont{and}
  \bibinfo{author}{\bibfnamefont{H.}~\bibnamefont{Hosono}},
  \bibinfo{journal}{J. Am. Chem. Soc.} \textbf{\bibinfo{volume}{130}},
  \bibinfo{pages}{3296} (\bibinfo{year}{2008}).

\bibitem[{\citenamefont{Chen et~al.}(2008{\natexlab{a}})\citenamefont{Chen, Li,
  Wu, Li, Hu, Dong, Zheng, Luo, and Wang}}]{ChenGF08prl}
\bibinfo{author}{\bibfnamefont{G.~F.} \bibnamefont{Chen}},
  \bibinfo{author}{\bibfnamefont{Z.}~\bibnamefont{Li}},
  \bibinfo{author}{\bibfnamefont{D.}~\bibnamefont{Wu}},
  \bibinfo{author}{\bibfnamefont{G.}~\bibnamefont{Li}},
  \bibinfo{author}{\bibfnamefont{W.~Z.} \bibnamefont{Hu}},
  \bibinfo{author}{\bibfnamefont{J.}~\bibnamefont{Dong}},
  \bibinfo{author}{\bibfnamefont{P.}~\bibnamefont{Zheng}},
  \bibinfo{author}{\bibfnamefont{J.~L.} \bibnamefont{Luo}}, \bibnamefont{and}
  \bibinfo{author}{\bibfnamefont{N.~L.} \bibnamefont{Wang}},
  \bibinfo{journal}{Phys. Rev. Lett.} \textbf{\bibinfo{volume}{100}},
  \bibinfo{pages}{247002} (\bibinfo{year}{2008}{\natexlab{a}}).

\bibitem[{\citenamefont{Chen et~al.}(2008{\natexlab{b}})\citenamefont{Chen, Wu,
  Wu, Liu, Chen, and Fang}}]{ChenXH08nature}
\bibinfo{author}{\bibfnamefont{X.~H.} \bibnamefont{Chen}},
  \bibinfo{author}{\bibfnamefont{T.}~\bibnamefont{Wu}},
  \bibinfo{author}{\bibfnamefont{G.}~\bibnamefont{Wu}},
  \bibinfo{author}{\bibfnamefont{R.~H.} \bibnamefont{Liu}},
  \bibinfo{author}{\bibfnamefont{H.}~\bibnamefont{Chen}}, \bibnamefont{and}
  \bibinfo{author}{\bibfnamefont{D.~F.} \bibnamefont{Fang}},
  \bibinfo{journal}{Nature} \textbf{\bibinfo{volume}{453}},
  \bibinfo{pages}{761} (\bibinfo{year}{2008}{\natexlab{b}}).

\bibitem[{\citenamefont{Mazin et~al.}(2008)\citenamefont{Mazin, Singh,
  Johannes, and Du}}]{Mazin08prl}
\bibinfo{author}{\bibfnamefont{I.~I.} \bibnamefont{Mazin}},
  \bibinfo{author}{\bibfnamefont{D.~J.} \bibnamefont{Singh}},
  \bibinfo{author}{\bibfnamefont{M.~D.} \bibnamefont{Johannes}},
  \bibnamefont{and} \bibinfo{author}{\bibfnamefont{M.~H.} \bibnamefont{Du}},
  \bibinfo{journal}{Phys. Rev. Lett.} \textbf{\bibinfo{volume}{101}},
  \bibinfo{pages}{057003} (\bibinfo{year}{2008}).

\bibitem[{\citenamefont{Kuroki et~al.}(2008)\citenamefont{Kuroki, Onari, Arita,
  Tanaka, Kontani, and Aoki}}]{Kuroki08prl}
\bibinfo{author}{\bibfnamefont{K.}~\bibnamefont{Kuroki}},
  \bibinfo{author}{\bibfnamefont{S.}~\bibnamefont{Onari}},
  \bibinfo{author}{\bibfnamefont{H.}~\bibnamefont{Arita}},
  \bibinfo{author}{\bibfnamefont{Y.}~\bibnamefont{Tanaka}},
  \bibinfo{author}{\bibfnamefont{H.}~\bibnamefont{Kontani}}, \bibnamefont{and}
  \bibinfo{author}{\bibfnamefont{H.}~\bibnamefont{Aoki}},
  \bibinfo{journal}{Phys. Rev. Lett.} \textbf{\bibinfo{volume}{101}},
  \bibinfo{pages}{087004} (\bibinfo{year}{2008}).

\bibitem[{\citenamefont{Bang and Choi}(2008)}]{Bang08prb}
\bibinfo{author}{\bibfnamefont{Y.}~\bibnamefont{Bang}} \bibnamefont{and}
  \bibinfo{author}{\bibfnamefont{H.-Y.} \bibnamefont{Choi}},
  \bibinfo{journal}{Phys. Rev. B} \textbf{\bibinfo{volume}{78}},
  \bibinfo{pages}{134523} (\bibinfo{year}{2008}).

\bibitem[{\citenamefont{Raghu et~al.}(2008)\citenamefont{Raghu, Qi, Liu,
  Scalapino, and Zhang}}]{Raghu08prb}
\bibinfo{author}{\bibfnamefont{S.}~\bibnamefont{Raghu}},
  \bibinfo{author}{\bibfnamefont{X.-L.} \bibnamefont{Qi}},
  \bibinfo{author}{\bibfnamefont{C.-X.} \bibnamefont{Liu}},
  \bibinfo{author}{\bibfnamefont{D.~J.} \bibnamefont{Scalapino}},
  \bibnamefont{and} \bibinfo{author}{\bibfnamefont{S.-C.} \bibnamefont{Zhang}},
  \bibinfo{journal}{Phys. Rev. B} \textbf{\bibinfo{volume}{77}},
  \bibinfo{pages}{220503} (\bibinfo{year}{2008}).

\bibitem[{\citenamefont{Chubukov et~al.}(2008)\citenamefont{Chubukov, Efremov,
  and Eremin}}]{Chubukov08prb}
\bibinfo{author}{\bibfnamefont{A.~V.} \bibnamefont{Chubukov}},
  \bibinfo{author}{\bibfnamefont{D.~V.} \bibnamefont{Efremov}},
  \bibnamefont{and} \bibinfo{author}{\bibfnamefont{I.}~\bibnamefont{Eremin}},
  \bibinfo{journal}{Phys. Rev. B} \textbf{\bibinfo{volume}{78}},
  \bibinfo{pages}{134512} (\bibinfo{year}{2008}).

\bibitem[{\citenamefont{Choi and Bang}(2008)}]{Choi0807.4604}
\bibinfo{author}{\bibfnamefont{H.~Y.} \bibnamefont{Choi}} \bibnamefont{and}
  \bibinfo{author}{\bibfnamefont{Y.}~\bibnamefont{Bang}},
  \bibinfo{journal}{arXiv:0807.4604}  (\bibinfo{year}{2008}).

\bibitem[{\citenamefont{Ding et~al.}(2008)\citenamefont{Ding, Richard,
  Nakayama, Sugawara, Arakane, Sekiba, Takayama, Souma, Sato, Takahashi
  et~al.}}]{Ding08epl}
\bibinfo{author}{\bibfnamefont{H.}~\bibnamefont{Ding}},
  \bibinfo{author}{\bibfnamefont{P.}~\bibnamefont{Richard}},
  \bibinfo{author}{\bibfnamefont{K.}~\bibnamefont{Nakayama}},
  \bibinfo{author}{\bibfnamefont{T.}~\bibnamefont{Sugawara}},
  \bibinfo{author}{\bibfnamefont{T.}~\bibnamefont{Arakane}},
  \bibinfo{author}{\bibfnamefont{Y.}~\bibnamefont{Sekiba}},
  \bibinfo{author}{\bibfnamefont{A.}~\bibnamefont{Takayama}},
  \bibinfo{author}{\bibfnamefont{S.}~\bibnamefont{Souma}},
  \bibinfo{author}{\bibfnamefont{T.}~\bibnamefont{Sato}},
  \bibinfo{author}{\bibfnamefont{T.}~\bibnamefont{Takahashi}},
  \bibnamefont{et~al.}, \bibinfo{journal}{Europhys. Lett.}
  \textbf{\bibinfo{volume}{83}}, \bibinfo{pages}{47001} (\bibinfo{year}{2008}).

\bibitem[{\citenamefont{Zhao}(2008)}]{Zhao0807.0398}
\bibinfo{author}{\bibfnamefont{L.}~\bibnamefont{Zhao}},
  \bibinfo{journal}{arXiv:0807.0398}  (\bibinfo{year}{2008}).

\bibitem[{\citenamefont{Kondo et~al.}(2008)\citenamefont{Kondo, Santander-Syro,
  Copie, Liu, Tillman, Mun, Schmalian, Bud'ko, Tanatar, Canfield
  et~al.}}]{Kondo0807.0815}
\bibinfo{author}{\bibfnamefont{T.}~\bibnamefont{Kondo}},
  \bibinfo{author}{\bibfnamefont{A.~F.} \bibnamefont{Santander-Syro}},
  \bibinfo{author}{\bibfnamefont{O.}~\bibnamefont{Copie}},
  \bibinfo{author}{\bibfnamefont{C.}~\bibnamefont{Liu}},
  \bibinfo{author}{\bibfnamefont{M.~E.} \bibnamefont{Tillman}},
  \bibinfo{author}{\bibfnamefont{E.~D.} \bibnamefont{Mun}},
  \bibinfo{author}{\bibfnamefont{J.}~\bibnamefont{Schmalian}},
  \bibinfo{author}{\bibfnamefont{S.~L.} \bibnamefont{Bud'ko}},
  \bibinfo{author}{\bibfnamefont{M.~A.} \bibnamefont{Tanatar}},
  \bibinfo{author}{\bibfnamefont{P.~C.} \bibnamefont{Canfield}},
  \bibnamefont{et~al.}, \bibinfo{journal}{arXiv:0807.0815}
  (\bibinfo{year}{2008}).

\bibitem[{\citenamefont{Matano et~al.}(2008)\citenamefont{Matano, Ren, Dong,
  Sun, Zhao, and q.~Zheng}}]{Matano0806.0249}
\bibinfo{author}{\bibfnamefont{K.}~\bibnamefont{Matano}},
  \bibinfo{author}{\bibfnamefont{Z.~A.} \bibnamefont{Ren}},
  \bibinfo{author}{\bibfnamefont{X.~L.} \bibnamefont{Dong}},
  \bibinfo{author}{\bibfnamefont{L.~L.} \bibnamefont{Sun}},
  \bibinfo{author}{\bibfnamefont{Z.~X.} \bibnamefont{Zhao}}, \bibnamefont{and}
  \bibinfo{author}{\bibfnamefont{G.}~\bibnamefont{q.~Zheng}},
  \bibinfo{journal}{arXiv:0806.0249}  (\bibinfo{year}{2008}).

\bibitem[{\citenamefont{Grafe et~al.}(2008)\citenamefont{Grafe, Paar, Lang,
  Curro, Behr, Werner, Hamann-Borrero, Hess, Leps, Klingeler
  et~al.}}]{Grafe0805.2595}
\bibinfo{author}{\bibfnamefont{H.-J.} \bibnamefont{Grafe}},
  \bibinfo{author}{\bibfnamefont{D.}~\bibnamefont{Paar}},
  \bibinfo{author}{\bibfnamefont{G.}~\bibnamefont{Lang}},
  \bibinfo{author}{\bibfnamefont{N.~J.} \bibnamefont{Curro}},
  \bibinfo{author}{\bibfnamefont{G.}~\bibnamefont{Behr}},
  \bibinfo{author}{\bibfnamefont{J.}~\bibnamefont{Werner}},
  \bibinfo{author}{\bibfnamefont{J.}~\bibnamefont{Hamann-Borrero}},
  \bibinfo{author}{\bibfnamefont{C.}~\bibnamefont{Hess}},
  \bibinfo{author}{\bibfnamefont{N.}~\bibnamefont{Leps}},
  \bibinfo{author}{\bibfnamefont{R.}~\bibnamefont{Klingeler}},
  \bibnamefont{et~al.}, \bibinfo{journal}{arXiv:0805.2595}
  (\bibinfo{year}{2008}).

\bibitem[{\citenamefont{Nakai et~al.}(2008)\citenamefont{Nakai, Ishida,
  Kamihara, Hirano, and Hosono}}]{Nakai0804.4765}
\bibinfo{author}{\bibfnamefont{Y.}~\bibnamefont{Nakai}},
  \bibinfo{author}{\bibfnamefont{K.}~\bibnamefont{Ishida}},
  \bibinfo{author}{\bibfnamefont{Y.}~\bibnamefont{Kamihara}},
  \bibinfo{author}{\bibfnamefont{M.}~\bibnamefont{Hirano}}, \bibnamefont{and}
  \bibinfo{author}{\bibfnamefont{H.}~\bibnamefont{Hosono}},
  \bibinfo{journal}{arXiv:0804.4765}  (\bibinfo{year}{2008}).

\bibitem[{\citenamefont{Fang et~al.}(2009)\citenamefont{Fang, Luo, Cheng, Wang,
  Jia, Mu, Shen, Mazin, Shan, Ren et~al.}}]{Fang0903.2418}
\bibinfo{author}{\bibfnamefont{L.}~\bibnamefont{Fang}},
  \bibinfo{author}{\bibfnamefont{H.}~\bibnamefont{Luo}},
  \bibinfo{author}{\bibfnamefont{P.}~\bibnamefont{Cheng}},
  \bibinfo{author}{\bibfnamefont{Z.}~\bibnamefont{Wang}},
  \bibinfo{author}{\bibfnamefont{Y.}~\bibnamefont{Jia}},
  \bibinfo{author}{\bibfnamefont{G.}~\bibnamefont{Mu}},
  \bibinfo{author}{\bibfnamefont{B.}~\bibnamefont{Shen}},
  \bibinfo{author}{\bibfnamefont{I.~I.} \bibnamefont{Mazin}},
  \bibinfo{author}{\bibfnamefont{L.}~\bibnamefont{Shan}},
  \bibinfo{author}{\bibfnamefont{C.}~\bibnamefont{Ren}}, \bibnamefont{et~al.},
  \bibinfo{journal}{arXiv:0903.2418}  (\bibinfo{year}{2009}).

\bibitem[{\citenamefont{Christianson et~al.}(2008)\citenamefont{Christianson,
  Goremychkin, Osborn, S.Rosenkranz, Lumsden, Malliakas, l.~S.~Todorov,
  H.Claus, Chung, Kanatzidis et~al.}}]{Christianson08nature}
\bibinfo{author}{\bibfnamefont{A.~D.} \bibnamefont{Christianson}},
  \bibinfo{author}{\bibfnamefont{E.~A.} \bibnamefont{Goremychkin}},
  \bibinfo{author}{\bibfnamefont{R.}~\bibnamefont{Osborn}},
  \bibinfo{author}{\bibnamefont{S.Rosenkranz}},
  \bibinfo{author}{\bibfnamefont{M.~D.} \bibnamefont{Lumsden}},
  \bibinfo{author}{\bibfnamefont{C.~D.} \bibnamefont{Malliakas}},
  \bibinfo{author}{\bibnamefont{l.~S.~Todorov}},
  \bibinfo{author}{\bibnamefont{H.Claus}},
  \bibinfo{author}{\bibfnamefont{D.~Y.} \bibnamefont{Chung}},
  \bibinfo{author}{\bibfnamefont{M.~G.} \bibnamefont{Kanatzidis}},
  \bibnamefont{et~al.}, \bibinfo{journal}{Nature}
  \textbf{\bibinfo{volume}{456}}, \bibinfo{pages}{930} (\bibinfo{year}{2008}).

\bibitem[{\citenamefont{Liu et~al.}(2008)\citenamefont{Liu, Wu, Wu, Chen, Wang,
  Xie, Yin, Yan, Li, Shi et~al.}}]{Liu0810.2694}
\bibinfo{author}{\bibfnamefont{R.~H.} \bibnamefont{Liu}},
  \bibinfo{author}{\bibfnamefont{T.}~\bibnamefont{Wu}},
  \bibinfo{author}{\bibfnamefont{G.}~\bibnamefont{Wu}},
  \bibinfo{author}{\bibfnamefont{H.}~\bibnamefont{Chen}},
  \bibinfo{author}{\bibfnamefont{X.~F.} \bibnamefont{Wang}},
  \bibinfo{author}{\bibfnamefont{Y.~L.} \bibnamefont{Xie}},
  \bibinfo{author}{\bibfnamefont{J.~J.} \bibnamefont{Yin}},
  \bibinfo{author}{\bibfnamefont{Y.~J.} \bibnamefont{Yan}},
  \bibinfo{author}{\bibfnamefont{Q.~J.} \bibnamefont{Li}},
  \bibinfo{author}{\bibfnamefont{B.~C.} \bibnamefont{Shi}},
  \bibnamefont{et~al.}, \bibinfo{journal}{arXiv:0810.2694}
  (\bibinfo{year}{2008}).

\bibitem[{\citenamefont{Boeri et~al.}(2008)\citenamefont{Boeri, Dolgov, and
  Golubov}}]{Boeri08prl}
\bibinfo{author}{\bibfnamefont{L.}~\bibnamefont{Boeri}},
  \bibinfo{author}{\bibfnamefont{O.~V.} \bibnamefont{Dolgov}},
  \bibnamefont{and} \bibinfo{author}{\bibfnamefont{A.~A.}
  \bibnamefont{Golubov}}, \bibinfo{journal}{Phys. Rev. Lett.}
  \textbf{\bibinfo{volume}{101}}, \bibinfo{pages}{026403}
  (\bibinfo{year}{2008}).

\bibitem[{\citenamefont{Franck}(1994)}]{Franck94review}
\bibinfo{author}{\bibfnamefont{J.~P.} \bibnamefont{Franck}},
  \emph{\bibinfo{title}{Physical Properties of High Temperature
  Superconductors, ed. D. M. Ginsberg}} (\bibinfo{publisher}{World Scientific},
  \bibinfo{address}{Singapore}, \bibinfo{year}{1994}), \bibinfo{note}{p. 189}.

\bibitem[{\citenamefont{Shimahara}(2003)}]{Shimahara03jpsj}
\bibinfo{author}{\bibfnamefont{H.}~\bibnamefont{Shimahara}},
  \bibinfo{journal}{J. Phys. Soc. Jpn.} \textbf{\bibinfo{volume}{72}},
  \bibinfo{pages}{1851} (\bibinfo{year}{2003}).

\bibitem[{\citenamefont{Dolgov et~al.}(2005)\citenamefont{Dolgov, Mazin,
  Golubov, Savrasov, and Maksimov}}]{Dolgov05prl}
\bibinfo{author}{\bibfnamefont{O.~V.} \bibnamefont{Dolgov}},
  \bibinfo{author}{\bibfnamefont{I.~I.} \bibnamefont{Mazin}},
  \bibinfo{author}{\bibfnamefont{A.~A.} \bibnamefont{Golubov}},
  \bibinfo{author}{\bibfnamefont{S.~Y.} \bibnamefont{Savrasov}},
  \bibnamefont{and} \bibinfo{author}{\bibfnamefont{E.~G.}
  \bibnamefont{Maksimov}}, \bibinfo{journal}{Phys. Rev. Lett.}
  \textbf{\bibinfo{volume}{95}}, \bibinfo{pages}{257003}
  (\bibinfo{year}{2005}).

\bibitem[{\citenamefont{Bickers et~al.}(1990)\citenamefont{Bickers, Scalapino,
  Collins, and Schlesinger}}]{Bickers00prb}
\bibinfo{author}{\bibfnamefont{N.~E.} \bibnamefont{Bickers}},
  \bibinfo{author}{\bibfnamefont{D.~J.} \bibnamefont{Scalapino}},
  \bibinfo{author}{\bibfnamefont{R.~T.} \bibnamefont{Collins}},
  \bibnamefont{and}
  \bibinfo{author}{\bibfnamefont{Z.}~\bibnamefont{Schlesinger}},
  \bibinfo{journal}{Phys. Rev. B} \textbf{\bibinfo{volume}{42}},
  \bibinfo{pages}{67} (\bibinfo{year}{1990}).

\bibitem[{\citenamefont{Bang}(2009)}]{Bang09prb}
\bibinfo{author}{\bibfnamefont{Y.}~\bibnamefont{Bang}}, \bibinfo{journal}{Phys.
  Rev. B} \textbf{\bibinfo{volume}{79}}, \bibinfo{pages}{092503}
  (\bibinfo{year}{2009}).

\end{thebibliography}

\end{document}